# Security Assessment of Software Design using Neural Network

A. Adebiyi, Johnnes Arreymbi and Chris Imafidon
School of Architecture, Computing and Engineering
University of East London,
London, UK

*Abstract*— Security flaws in software applications today has been attributed mostly to design flaws. With limited budget and time to release software into the market, many developers often consider security as an afterthought. Previous research shows that integrating security into software applications at a later stage of software development lifecycle (SDLC) has been found to be more costly than when it is integrated during the early stages. To assist in the integration of security early in the SDLC stages, a new approach for assessing security during the design phase by neural network is investigated in this paper. Our findings show that by training a back propagation neural network to identify attack patterns, possible attacks can be identified from design scenarios presented to it. The result of performance of the neural network is presented in this paper.

I. INTRODUCTION

The role software applications play in today's hostile computer environment is very important. It is not uncommon to find software applications running our transportation systems, communication systems, medical equipment, banking systems, domestic appliances and other technologies that we depend on. Since many of these software applications are missions critical, the need to ensure the security of their data and other resources cannot be overlooked. The increase of attacks aimed directly at software applications in the past decades calls for software applications to be able to defend itself and continue functioning. However, when software applications are developed without security in mind, attackers take advantage of the security flaws in them to mount multiple attacks when they are deployed. To address this problem a new research field called software security emerged in the last decade with the aim of building security into software application during development. This approach views security as an emergent property of the software and much effort is dedicated into weaving security into the software all through SDLC

One of the critical areas in this approach is the area of software design and security which proactively deals with attacking security problems at the design phase of SDLC. Reportedly, 50% of security problems in software products today have been found to be design flaws [17]. Design-level vulnerability has been described as the hardest category of software defect to contend with. Moreover, it requires great expertise to ascertain whether or not a software application has design-level flaws which makes it difficult to find and automate [9]. Many authors also argue that it is much better to find and fix flaws during the early phase of software development because it is more costly to fix the problem at a late stage of development and much more costly when the software has been deployed [6][29][30]. Therefore, taking security into consideration at the design phase of SDLC will help greatly in producing secured software applications.

There are different approaches and tools currently used for integrating security during the phases of SDLC. However, software design security tools and technologies for automated security analysis at the design phase have been slow in coming. This is still an area where many researches are currently being undertaken. Neural Networks has been one of the technologies used during software implementation and testing phase of SDLC for software defect detection in order to intensify software reliability and it has also been used in area of application security and network security in technologies such as authentication system, cryptography, virus detection system, misuse detection system and intrusion detection systems (IDS) [2] [4] [14] [20] [31][32]. This research takes a further step by using neural networks as a tool for assessing security of software design at the design phase of SDLC.

II. RELATED WORKS ON SECURITY ASSESSMENT OF SOFTWARE DESIGN

In order to design software more securely many approaches have been adopted for assessing the security in software designs during the design phase of SDLC. Some of these approaches are discussed below.

Threat modeling is an important activity carried out at the design phase to describe threats to the software application in order to provide a more accurate sense of its security [1]. Threat modeling is a technique that can be used to identify vulnerabilities, threats, attacks and countermeasures which could influence a software system [18]. This allows for the anticipation of attacks by understanding how a malicious attacker chooses targets, locates entry points and conducts attacks [24]. Threat modeling addresses threats that have the ability to cause maximum damage to a software application.

Architectural risk analysis is also used to identify vulnerabilities and threats at the design phase of SDLC which may be malicious or non-malicious in nature due to a software system. It examines the preconditions that must be present for the vulnerabilities to be exploited by various threats and assess the states the system may enter after a successful attack on the system. One of the advantages of architectural risk analysis is that it enables developers to analysis software system from its component level to its environmental level in order to evaluate the vulnerabilities, threats and impacts at each level [17].

Attack trees is another approach used to characterize system security by modeling the decision making process of attackers. In this technique, attack against a system is represented in a tree structure in which the root of the tree represents the goal of an attacker. The nodes in the tree represent the different types of actions the attacker can take to accomplish his goal on the software system or outside the software system which may be in the form of bribe or threat [6],[23]. "Attack trees are used for risk analysis, to answer questions about the system's security, to capture security knowledge in a reusable way, and to design, implement, and test countermeasures to attacks" [24].

Attack nets is a similar approach which include "places" analogous to the nodes in an attack tree to indicate the state of an attack. Events required to move from one place to the other are captured in the transitions and arcs connecting places and transitions indicate the path an attacker takes. Therefore just as attack trees, attack nets also show possible attack scenarios to a software system and they are used for vulnerability assessment in software designs [6].

Another related approach is the vulnerability tree which is a hierarchy tree constructed based on how one vulnerability relates to another and the steps an attacker has to take to reach the top of the tree [23]. Vulnerability trees also help in the analysis of different possible attack scenarios that an attacker can undertake to exploit a vulnerability.

Gegick and Williams [6] also proposed a regular expression-based attack patterns which helps in indicating the sequential events that occur during an attack. The attack patterns are based on the software components involved in an attack and are used for identifying vulnerabilities in software designs. It comprises of attack library of abstraction which can be used by software engineers conducting Security Analysis For Existing Threats (SAFE-T) to match their system design. An occurrence of a match indicates that the vulnerability may exist in the system being analyzed and therefore helps in integrating effective countermeasures before coding starts. Another advantage about this approach is that it can be easily adapted by developers who are novices on security.

Mouratidis and Giorgini [19] also propose a scenario based approach called Security Attack Testing (SAT) for testing the security of a software system during design time. To achieve this, two sets of scenarios (dependency and security attack) are identified and constructed. Security test cases are then defined from the scenarios to test the software design against potential attacks to the software system.

Essentially SAT is used to identify the goals and intention of possible attackers based on possible attack scenarios to the system. Software engineers are able to evaluate their software design when the attack scenarios identified are applied to investigate how the system developed will behave when under such attacks. From this, software engineers better understand how the system can be attacked and also why an attacker may want to attack the system. Armed with this knowledge, necessary steps can be taken to secure the software with capabilities that will help in mitigating such attacks

For most of the approaches discussed above, the need to involve security experts is required in order to help in identifying the threats to the software technology, review the software for any security issues, investigate how easy it is to compromise the software's security, analysis the impact on assets and business goals should the security of the software be compromised and recommend mitigating measures to either eliminate the risk identified or reduce it to a minimum. The need for security experts arises because there is an existing gap between security professionals and software developers. The disconnection between this two has led to software development efforts lacking critical understanding of current technical security risks [22].

In a different approach, Kim T. et.al [12] introduced the notion of dynamic software architecture slicing (DSAS) through which software architecture can be analyzed. "A dynamic software architecture slice represents the run-time behavior of those parts of the software architecture that are selected according to a particular slicing criterion such as a set of resources and events" [12] DSAS is used to decompose software architecture based on a slicing criterion. "A slicing criterion provides the basic information such as the initial values and conditions for the ADL (Architecture description language) executable, an event to be observed, and occurrence counter of the event" [12] While software engineers are able to examine the behavior of parts of their software architecture during run time using the DSAS approach, the trade-off is that it requires the software to be implemented first. The events examined to compute the architecture slice dynamically are generated when the Forward Dynamic Slicer executes the ADL executable. This is a drawback because fixing the vulnerability after implementation can be more costly [6].

Howe [10] also argues that the industry needs to invest in solutions that apply formal methods in analyzing software specification and design in order to reduce the number of defects before implementation starts. "Formal methods are mathematically based techniques for the specification development and verification of software and hardware systems" [7] Recent advances in formal methods have also made verification of memory safety of concurrent systems possible [7].

As a result, formal methods are being used to detect design errors relating to concurrency [10]. A software development process incorporating formal methods into the overall process of early verification and defects removal through all SDLC is Correct by Construction (CbyC) [24]. CbyC has proved to be very cost effective in developing software because errors are eliminated early during SDLC or not introduced in the first place. This subsequently reduces the amount of rework that would be needed later during software development. However, many software development organizations have been reluctant in using formal methods because they are not used to its rigorous mathematical approach in resolving security issues in software design. Model checkers also come with their own modeling language which makes no provision for automatically translating informal requirements to this language. Therefore, the translation has to be done manually and it may be difficult to check whether the model represent the target system [21]

### III. THE NUERAL NETWORK APPROACH

Our proposed Neural Network approach in analysing software design for security flaws is based on the abstract and match technique through which software flaws in a software design can be identified when an attack pattern is matched to the design. Using the regularly expressed attack patterns proposed by Williams and Gegick [6], the actors and software components in each attack pattern are identified. To generate the attack scenarios linking the software components and actors identified in the attack pattern, online vulnerability databases were used to identify attack scenarios corresponding to the attack pattern. Data of attack scenarios from online vulnerability databases such as CVE Details, Security Tracker, Secunia, Security Focus and The Open Source Vulnerability Database were used.

*A. The Neural Network Architecture*

A three-layered feed-forward back-propagation was chosen for the architecture of neural network in this research. The back-propagation neural network is a well-known type of neural network commonly used in pattern recognition problems [25]. A back-propagation network has been used because of its simplicity and reasonable speed.

The architecture of the neural network consists of the input layer, the hidden layer and the output layer. Each of the hidden nodes and output nodes apply a tan-sigmoid transfer function $(2/(1+exp(-2*n))-1)$ to the various connection weights.

The weights and parameters are computed by calculating the error between the actual and expected output data of the neural network when the training data is presented to it. The error is then used to modify the weights and parameters to enable the neural network to a have better chance of giving a correct output when it is next presented with same input

*B. Data Collection*

From the online vulnerability databases mentioned above, a total of 715 attack scenarios relating to 51 regularly expressed attack patterns by Williams and Gegick's were analysed. This consisted of 260 attack scenarios which were unique in terms of their impact, mode of attack, software component and actors involved in the attack and 455 attack scenarios which are repetition of the same type of exploit in different applications they have been reported in the vulnerability databases. The attacks were analysed to identify the actors, goals and resources under attack.

Once these were identified the attack attributes below were used to abstract the data capturing the attack scenario for training the neural network. The attack attributes includes the following.

1. The Attacker: This attribute captures the capability of the attacker. It examines what level of access possessed when carrying out the attack.
2. Source of attack: This attributes captures the location of the attack during the attack.
3. Target of the attack: This captures the system component that is targeted by the attacker
4. Attack vector: This attributes captures the mechanism (i.e. software component) adopted by the attacker to carry out the attack
5. Attack type: The security property of the application being attacked is captured under this attribute. This could be confidentiality, integrity or availability.
6. Input Validation: This attributes examines whether any validation is done on the input passed to the targeted software application before it is processed
7. Dependencies: The interaction of the targeted software application with the users and other systems is analysed under this attributes.
8. Output encoding to external applications/services: Software design scenarios are examined under this attributes to identify attacks associated with flaws due to failure of the targeted software application in properly verifying and encoding its outputs to other software systems
9. Authentication: This attribute checks for failure of the targeted software application to properly handle account credentials safely or when the authentication is not enforced in the software design scenarios.
10. Access Control: Failure in enforcing access control by the targeted software application is examined in the design scenarios with this attribute.
11. HTTP Security: Attack Scenarios are examined for security flaws related to HTTP requests, headers, responses, cookies, logging and sessions with this attribute
12. Error handling and logging: Attack scenarios are examined under this attributes for failure of the targeted application in safely handling error and security flaws in log management.

*C. Data Encoding*

The training data samples each consist of 12 input units for the neural network. This corresponds to the values of the attributes abstracted from the attack scenarios.

The training data was generated from the attack scenarios using the attributes. For instance training data for the attack on webmail (CVE 2003-1192) was generated by looking at the online vulnerability databases to get its details on the attributes we are interested in.

This attack corresponds to regularly expressed attack pattern 3. Williams and Gegick [6] describe the attack scenario in this attack pattern as a user submitting an excessively long HTTP GET request to a web server, thereby causing a buffer. This attack pattern is represented as:

(User)(HTTPServer)(GetMethod)(GetMethodBufferWrite)(Buffer)

TABLE I. SAMPLE OF PRE-PROCESSED TRAINING DATA FROM ATTACK SCENARIO

| S\N | Attribute | Observed data |
|---|---|---|
| 1 | Attacker | No Access |
| 2 | Source | External |
| 3 | Target | Buffer |
| 4 | Attack Vector | Long Get Request |
| 5 | Attack Type | Availability |
| 6 | Input Validation | Partial Validation |
| 7 | Dependencies | Authentication & Input Validation |
| 8 | Output Encoding | None |
| 9 | Authentication | None |
| 10 | Access Control | URL Access |
| 11 | HTTP Security | Input Validation |
| 12 | Error | None |

In this example, the data generated from the attack scenario using the attribute list is shown in Table I. Using the corresponding values for the attributes; the data is then encoded as shown in the

TABLE II. SAMPLE OF TRAINING DATA AFTER ENCODING

| S\N | Attribute | Value |
|---|---|---|
| 1 | Attacker | 0 |
| 2 | Source | 1 |
| 3 | Target | 9 |
| 4 | Attack Vector | 39 |
| 5 | Attack Type | 5 |
| 6 | Input Validation | 2 |
| 7 | Dependencies | 6 |
| 8 | Output Encoding | 0 |
| 9 | Authentication | 0 |
| 10 | Access Control | 2 |
| 11 | HTTP Security | 3 |
| 12 | Error | 0 |

The second stage of the data processing involves converting the value of the attributes in Table II into ASCII comma delimited format before it is used in training the neural network. For the expected output from the neural network, the data used in training network is derived from the attack pattern which has been identified in each of the attack scenarios. Each attack pattern is given a unique ID which the neural network is expected to produce as an output for each of the input data samples. The output data sample consists of output units corresponding to the attack pattern IDs. For instance, the above sample data on Webmail attack which corresponds to regularly expressed attack pattern 3, the neural network is trained to identify the expected attack pattern as 3.

### D. The Neural Network Training

To train the neural network the training data set is divided into two sets. The first set of data is the training data sets (260 samples) that were presented to the neural network during training.

TABLE III. TRAINING AND TEST DATA SETS

| Number of Samples | Training Data | Test Data |
|---|---|---|
| Data Set 1 | 143 | 26 |
| Data set 2 | 117 | 25 |
| Total | 260 | 51 |

The second set (51 Samples) is the data that were used to test the performance of the neural network after it had been trained. At the initial stage of the training, it was discovered that the neural network had too many categories to classify the input data into (i.e. 51 categories) because the neural network was not able to converge. To overcome the problem, the training data was further divided into two sets. The first set contained 143 samples and the second set contained 117 samples. These were then used for training two neural networks. Mat lab Neural Network tool box is used to perform the training. The training performance is measured by Mean Squared Error (MSE) and the training stops when the generalization stops improving or when the 1000th iteration is reached.

### E. Result and Discussion

It took the system about one minute to complete the training for each the back-propagation neural network. For the first neural network, the training stopped when the MSE of 0.0016138 was reached at the 26th iteration. The training of the second neural network stopped when the MSE of 0.00012841 was reached at the 435[th] iteration.

TABLE IV. COMPARISION OF ACTUAL AND EXPECTED OUTPUT FROM NEURAL NETWORK

| s\n | Attack Pattern Investigated | Actual Output | Expected Output |
|---|---|---|---|
| | Results from Neural Network 1 | | |
| 1 | Attack Pattern 1 | 1.0000 | 1 |
| 2 | Attack Pattern 2 | 2.0000 | 2 |
| 3 | Attack Pattern 3 | 2.9761 | 3 |
| 4 | Attack Pattern 4 | 4.0000 | 4 |
| 5 | Attack Pattern 5 | 4.9997 | 5 |
| 6 | Attack Pattern 6 | 5.9998 | 6 |
| 7 | Attack Pattern 7 | 7.0000 | 7 |
| 8 | Attack Pattern 8 | 8.0000 | 8 |
| 9 | Attack Pattern 9 | 9.0000 | 9 |
| 10 | Attack Pattern 10 | 7.0000 | 10 |
| 11 | Attack Pattern 11 | 11.0000 | 11 |
| 12 | Attack Pattern 12 | 12.0000 | 12 |
| 13 | Attack Pattern 13 | 12.9974 | 13 |
| 14 | Attack Pattern 14 | 13.772 | 14 |
| 15 | Attack Pattern 15 | 15.0000 | 15 |
| 16 | Attack Pattern 16 | 16.0000 | 16 |
| 17 | Attack Pattern 17 | 16.9999 | 17 |
| 18 | Attack Pattern 20 | 19.9984 | 20 |
| 19 | Attack Pattern 21 | 21.0000 | 21 |
| 20 | Attack Pattern 22 | 22.0000 | 22 |
| 21 | Attack Pattern 23 | 23.0000 | 23 |
| 22 | Attack Pattern 24 | 23.9907 | 24 |
| 23 | Attack Pattern 25 | 25.0000 | 25 |
| 24 | Attack Pattern 26 | 26.0000 | 26 |
| 25 | Attack Pattern 27 | 27.0000 | 27 |
| 26 | Attack Pattern 28 | 28.0000 | 28 |
| Results from Network 2 | | | |
| 27 | Attack Pattern 29 | 28.999 | 29 |
| 28 | Attack Pattern 30 | 29.9983 | 30 |
| 29 | Attack Pattern 31 | 31.0000 | 31 |
| 30 | Attack Pattern 32 | 31.998 | 32 |
| 31 | Attack Pattern 33 | 32.8828 | 33 |
| 32 | Attack Pattern 34 | 33.9984 | 34 |
| 33 | Attack Pattern 35 | 32.8828 | 35 |
| 34 | Attack Pattern 36 | 35.9945 | 36 |
| 35 | Attack Pattern 37 | 36.6393 | 37 |
| 36 | Attack Pattern 38 | 37.9999 | 38 |
| 37 | Attack Pattern 39 | 37.9951 | 39 |
| 38 | Attack Pattern 40 | 39.1652 | 40 |
| 39 | Attack Pattern 41 | 40.9669 | 41 |
| 40 | Attack Pattern 42 | 41.9998 | 42 |
| 41 | Attack Pattern 43 | 42.998 | 43 |
| 42 | Attack Pattern 44 | 43.9979 | 44 |
| 43 | Attack Pattern 45 | 44.9991 | 45 |
| 44 | Attack Pattern 46 | 45.8992 | 46 |
| 45 | Attack Pattern 47 | 46.9956 | 47 |
| 46 | Attack Pattern 48 | 47.9997 | 48 |
| 47 | Attack Pattern 49 | 48.9999 | 49 |
| 48 | Attack Pattern 50 | 49.8649 | 50 |
| 49 | Attack Pattern 51 | 50.9629 | 51 |
| 50 | Attack Pattern 52 | 50.6745 | 52 |
| 51 | Attack Pattern 53 | 52.7173 | 53 |

To test the performance of the network, the second data sets were used to test the neural network. It was observed that the trained neural network gave an output as close as possible to the anticipated output. The actual and anticipated outputs are compared in the Table IV. The test samples in which the neural network gave a different output from the predicted output when testing the network includes tests for attack patterns 10, 35, 39, 40 and 52. While looking into the reason behind this, it was seen that the data observed for these attack patterns were not much. With more information on these attack patterns for training the neural network, it is predicted that the network will give a better performance. During the study of the results from the neural networks, it was found that the first neural network had 96.51% correct results while the second neural network had 92% accuracy. The accuracy for both neural networks had an average of 94.1%. Given the accuracy of the neural networks, it shows that neural networks can be used to assess the security in software designs.

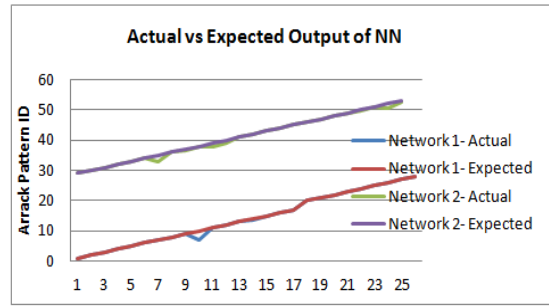

Figure 1. Actual vs. Expected output of Neural Network

### IV. FUTURE WORK

To further improve the performance of the neural network system as a tool for assessing security in software designs, we are currently looking into the possibility of the system suggesting solutions that can help to prevent the identified attacks. Current research on solutions to software design security flaws gives a good insight in this area. Suggested solutions such as the use security patterns [11] and introduction of security capabilities into design in the SAT approach [19] are currently investigated. Furthermore, the performance of the neural network tool would be compared to current approaches used in assessing security in software designs in a case study on the design of an online shopping portal.

The regularly expressed attack pattern used in training the neural network is a generic classification of attack patterns Therefore; any unknown attack introduced to the neural network will be classified to the nearest regularly expressed attack pattern. Nevertheless the success of the neural network in analysing software design for security flaws is largely dependent upon the input capturing the features of the software design presented to it. As this requires a human endeavour, further work is required in this area to ensure that correct input data is retrieved for analysis. In addition, the neural network needs to be thoroughly tested before it can gain acceptance as a tool for assessing software design for security flaws.

### V. CONCLUSION

Previous research works have shown that the cost of fixing security flaws in software applications when they are deployed is 4–8 times more than when they are discovered early in the SDLC and fixed. For instance, it is cheaper and less disruptive to discover design-level vulnerabilities in the design, than during implementation or testing, forcing a pricey redesign of pieces of the application. Therefore, integrating security into a software design will help tremendously in saving time and money during software development

Therefore, by using the proposed neural networks approach in this paper to analyse software design for security flaws the efforts of software designers in identifying areas of security weakness in their software design will be reinforced. Subsequently, this will enhance the development of secured software applications in the software industry especially as software designers often lack the required security expertise. Thus, neural networks given the right information for its training will also contribute in equipping software developers to develop software more securely especially in the area of software design.

AUTHORS PROFILE

**Adetunji Adebiyi** Doctoral student with the University of East London UK. His research focuses on integrating security into software design during SDLC. His research has led him to give talks and presentations in conferences and seminars he has attended.

**Johnnes Arreymbi** is a Senior Lecturer at the School of Computing, Information Technology and Engineering, University of East London. He has also taught Computing at London South Bank University, and University of Greenwich, London. He leads as Executive Director and co-founder of eGLobalSOFT, USA; an innovative Patented Software (ProTrack™) Company.

**Chris Imafidon** is a Senior Lecturer at the School of Computing, Information Technology and Engineering, University of East London. Chris is a multi-award winning researcher and scientific pioneer. He is a member of the Information Age Executive Round-table forum. He is a consultant to the government and industry leaders worldwide.